\documentclass[structabstract]{aa}  
\usepackage{graphicx}
\usepackage{natbib,twoopt}
\usepackage[breaklinks=true]{hyperref} 

\bibpunct{(}{)}{;}{a}{}{,}    
\newcommandtwoopt{\citeads}[3][][]{\href{http://adsabs.harvard.edu/abs/#3}%
                                        {\citealp[#1][#2]{#3}}}
\newcommandtwoopt{\citepads}[3][][]{\href{http://adsabs.harvard.edu/abs/#3}%
                                        {\citep[#1][#2]{#3}}}
\newcommandtwoopt{\citetads}[3][][]{\href{http://adsabs.harvard.edu/abs/#3}%
                                        {\citet[#1][#2]{#3}}}
\newcommandtwoopt{\citeyearads}[3][][]%
   {\href{http://adsabs.harvard.edu/abs/#3}{\citeyear[#1][#2]{#3}}}
\usepackage{txfonts}
\begin{document}
\title{Linear polarization of submillimetre masers}

   \subtitle{Tracing magnetic fields with ALMA}

\authorrunning{P\'erez-S\'anchez \and
               Vlemmings}

   \author{A. F. P\'erez-S\'anchez
          \inst{1}
          \and
          W. H. T. Vlemmings\inst{2}
          }

   \institute{Argelander Institute for Astronomy, University of Bonn,
              Auf dem H\"ugel 71, 53121 Bonn, Germany\\
              \email{aperez@astro.uni-bonn.de}
         \and
             Chalmers University of Technology, Onsala Space Observatory,
             SE-439 92 Onsala, Sweden.\\
             \email{wouter.vlemmings@chalmers.se}
             }

   \date{}


  \abstract
     {Once ALMA full polarization capabilities are offered, it will become possible to perform detailed studies of polarized maser emission towards
    star-forming regions and late-type stars, such as (post-) asymptotic giant branch stars and young planetary nebulae.
    To derive the magnetic field orientation from maser linear polarization, a number of conditions involving the rate of stimulated emission $R$, the decay 
    rate of the molecular state $\Gamma$, and the Zeeman frequency $g\Omega$ need to
    be satisfied.}
     {The goal of this work is to investigate if SiO, H$_{2}$O and HCN maser emission within the ALMA frequency range can be detected with 
    observable levels of fractional linear polarization in the regime where the Zeeman frequency is greater than the stimulated emission rate.}
     {We used a radiative transfer code to calculate the fractional linear polarization as a function of the emerging brightness temperature 
      for a number of rotational transition of SiO, H$_{2}$O and HCN that have been observed to display maser emission at submillimetre
     wavelengths. We assumed typical magnetic field strengths measured towards galactic star-forming regions and circumstellar envelopes of late-type stars
     from previous VLBI observations. Since the Land\'e g-factors have not been reported for the different rotational transitions we modelled, 
     we performed our calculations assuming conservative values of the Zeeman frequency for the different molecular species.}
     {Setting a lower limit for the Zeeman frequency that still satisfies the criteria $g\Omega>R$ and $g\Omega>\Gamma$, we find 
    fractional polarization levels of up to 13\%, 14\% and 19\% for the higher $J$ transitions analysed for SiO, H$_{2}$O and HCN, respectively, without
    considering anisotropic pumping or any other non-Zeeman effect. These upper limits were calculated assuming a magnetic field oriented perpendicular to 
    the direction of propagation of the maser radiation.} 
    {According to our results SiO, H$_{2}$O and HCN maser emission within the ALMA frequency range can be detected with suitable linear polarization to 
    trace the magnetic field structure towards star-forming regions and late-type stars even if the detected polarization has been enhanced by non-Zeeman
    effects.}


   \maketitle
%

\section{Introduction}
Polarized maser emission has been detected towards star-forming
regions (SFR) and expanding circumstellar envelopes (CSE) of late-type stars such as
(post-) asymptotic giant branch (AGB) stars and young planetary nebulae 
(PNe) (e.g. \citealp{Alves, Amiri, Ferreira, Vlemmings11, Vlemmings1}). Both single-dish and interferometric observations have revealed that 
silicon monoxide (SiO), water (H$_{2}$O), 
hydrogen cyanide (HCN), hydroxyl (OH), and methanol (CH$_{3}$OH), among others, can naturally generate polarized maser emission 
in these enviroments (e.g. \citealp{Vlemmings1, Surcis, Fish, Herpin, Kemball09}).
In particular, interferometric observations of the masers at radio wavelengths have become a useful tool for studying the magnetic field in 
and around SFRs and late-type stars (e.g. \citealp{Fish07,Amiri10}). 
At shorter wavelengths, new instruments will
enable the study of maser radiation from higher vibrationally-excited rotational transitions. In particular, the Atacama Large
Millimeter/submillimeter Array (ALMA) has recently started the first scientific observations. Soon all of its capabilities, including polarimetry, will be 
available, providing more than an order of magnitude improvement in
sensitivity and resolution. In the ALMA frequency range, a number of SiO, H$_{2}$O and HCN maser 
transitions that belong to vibrationally-excited levels up to $v=3$ can be detected. The population inversion
of rotational states belonging to different molecular species, or within different vibrational levels of a particular species,
depends on the physical conditions of the emitting regions. By observing the maser polarization at different frequencies towards
an extended source, we can hence constrain the magnetic field properties, i.e. field strength and/or direction, throughout large regions within the observed source.

According to the Zeeman interpretation, the maser polarization strongly depends on the ratio between the Zeeman 
frequency ($g\Omega$), the rate of stimulated emission ($R$), and the rate of the decay of the molecular 
state ($\Gamma$). The Zeeman splitting induced by a magnetic field depends on the molecule's shell structure.
For example, since SiO, H$_{2}$O and HCN are non-paramagnetic, closed-shell molecules, $g\Omega$ is expected to be less 
than the intrinsic line width $\Delta\omega$. 
Maser polarization theory (e.g \citealt{Western}) predicts fractional linear polarization ($p_{L}$) levels of up to 100$\%$ 
for $J=1-0$ rotational transitions of diatomic molecules. These polarization levels can be reached in the 
presence of a magnetic field of a few Gauss in, for example, the SiO masing region in CSEs of late-type stars. 
But in the case of higher rotational transitions (i.e. $J=2-1$, $J=3-2$, etc.) theory predicts that unless anisotropic
pumping is involved, the fractional linear polarization should decrease as the angular momentum number of the involved state increases.

Before the observed maser polarization can be related to an intrinsic magnetic
field, it is necessary to evaluate the ratios between the maser parameters $R$, $g\Omega$ and $\Gamma$ for each single rotational transition 
detected with angular momentum higher than $J=1$. Here, we determine if the polarized maser radiation produced
by molecular transitions of SiO, H$_{2}$O and HCN in the ALMA frequency range could generate detectable levels of
fractional linear polarization, while still satisfying the criteria
($g\Omega>R$ and $g\Omega>\Gamma$) for which the polarization
direction is still directly related to the magnetic field. These
maser transitions can then be used to infer the magnetic field 
structure towards SFRs and in the CSEs of (post-) AGB stars. 
To do this, we have run numerical models adapted from \citet{NW92} to calculate
the fractional linear polarization level that can be generated by the interaction of the molecular states of SiO, H$_{2}$O 
and HCN in the ALMA frequency range, with a magnetic field in the masing region.

In section 2 we present a brief review of the maser polarization
theory and in section 3 we describe the different maser
transitions of SiO, H$_{2}$O and HCN that can be detected in the ALMA frequency range. The results of our models
for each molecular species are detailed in section 4. In section 5
we give an estimate of the polarized fluxes that can be expected when observing with ALMA.   

\section{Maser polarization theory}

Polarized maser radiation provides a unique tool for studying the role of magnetic fields inside
high-density enviroments such as star-forming regions and circumstellar envelopes of evolved stars.
Although the general maser emission mechanims from astronomical sources have been well understood throughout the past thirty years
(e.g. \citealt{ElitzurB}, \citealt{grayB}), the generation and radiative transfer of polarized
maser emission have been difficult to deal with. The polarization properties
of maser emission strongly depend on the radiative conditions of the region where 
the maser is being generated (saturated or unsaturated) and on the nature of the molecular species generating the 
maser emission (paramagnetic or non-paramagnetic) \citep{Watson08, Dinh}.\\
Maser-polarized radiation can be produced in both unsaturated and saturated frames.
Maser emission is considered to be saturated when the rate for the stimulated emission $R$ overcomes the decay rate of the molecular state 
involved, $\Gamma$. In this case, the growth of the possible polarization modes is determined by 
the population of the molecular states that can interact with a particular mode of the polarization \citep{NW90}.
\citet{Goldreich} first identified two regimes where polarized emission can be generated in the saturated frame
for the molecular rotational transition $J=1-0$: a) The strong magnetic
field strength regime, where the Zeeman frequency
$g\Omega\gg\ R$ and $g\Omega\gg\ \Gamma$; and b) The intermediate magnetic field strength regime, where ($g\Omega)^{2}/\Gamma\gg\ R\ \gg\ g\Omega$. 
\citet{Western}, \citet{Deguchi} and \citet{NW90} extended the treatment
of \citet{Goldreich}, solving the radiative transfer equations for polarized radiation as a function of 
the emerging intensity $R/\Gamma$, i.e. as a function of the saturation level. They considered linear, non-paramagnetic molecules, including the rotational 
transitions $J=2-1$ and $J=3-2$. \citet{NW90} have shown that in the presence of a plausible magnetic field and
without differences in the population of the magnetic substates involved in the maser emission, the
upper limit of fractional linear polarization $p_{L}$ that maser transitions other than $J=1-0$ can achieve is $33\%$. The fractional linear polarization 
decreases for transitions with higher angular momentum, and is a function of the angle $\theta$ between 
the magnetic field (\vec{B}) and the direction of the maser radiation (\vec{k}).\\
Molecular states with angular momentum higher than $J=1-0$ reach their maximum $p_{L}$ value when the magnetic field lines are perpendicular to 
the direction of the propagation of the maser radiation, and
have $p_{L}=0$ when $\theta$ has the critical value 
$\theta_{cr}\sim 55^{\circ}$, also known as the van Vleck angle, and when $\theta=0^\circ$.\\
The angle $\phi$ between the polarization vector and the plane $\vec{k}\cdot\vec{B}$ 
also varies as a function of the ratio $g\Omega$/$R$. For a fixed $\theta$, the vector of 
polarization should be either perpendicular or parallel to the $\vec{k}\cdot\vec{B}$ plane if the condition $g\Omega\gg$ $R$ is satisfied. 
On the other hand, for higher values of $R$, but still in the intermediate-strength magnetic field regime, the vector
of polarization is already rotated away from the $\vec{k}\cdot\vec{B}$ plane, i.e.~if $R \ >\ g\Omega$ 
the vector of polarization is neither parallel nor perpendicular to the plane $\vec{k}\cdot\vec{B}$. In this context, 
the van Vleck angle is the limit for $\theta$ where the polarization vector changes from being parallel 
to be perpendicular to the $\vec{k}\cdot \vec{B}$ plane. Therefore, as long as the conditions $g\Omega > R$ and $g\Omega > \Gamma$ are satisfied, 
the information about the morphology of the magnetic field can be extracted from the polarization vector.

Observationally, fractional linear polarization of up to 100$\%$ for $J=1-0$ SiO maser transitions \citep{Amiri}
and values higher than 33$\%$ for SiO molecular transitions involving higher angular momentum states 
(e.g \citealt{Vlemmings11}) have been detected. Anisotropic pumping seems to be a possible process to 
explain such high $p_{L}$ levels \citep{NW90}. It can be produced by differences in the angular
distribution of the radiation field involved in the population inversion process. In the absence of a magnetic
field, anisotropic pumping could produce highly linearly polarized emission. In contrast to the effect of a magnetic 
field alone in the maser region, the fractional linear polarization can increase with the angular momentum due to 
an anisotropic population of the involved magnetic substates. Nevertheless, even considering that 
the linear polarization has been enhanced by anisotropic pumping, if $g\Omega > R$ and $g\Omega >\Gamma$ for the detected line, 
the magnetic field structure can still be traced directly from the polarization vector, because the magnetic field is the dominant axis 
of symmetry \citep{Watson}. The study of magnetic fields from maser linear polarization thus depends critically on an analysis of the Zeeman frequency $g\Omega$
and the stimulated emission rate $R$.

\begin{table*}
\caption{Spontaneous emission coefficient and frequencies of the detected SiO maser transitions in the ALMA frequency range}             
\label{table:1}      
\centering                          
\begin{tabular}{c c c c c | c c c c c}
\hline\hline                 
\multicolumn{5}{c}{$^{28}$SiO} & \multicolumn{5}{c}{$^{29}$SiO}\\
\hline
$v$ & $J_{u}-J_{d}$  & Freq  & A & ALMA & $v$ & $J_{u}-J_{d}$ & Freq & A & ALMA\\    
      &   & (GHz) & (s$^{-1}$) & band & & & (GHz) & (s$^{-1}$) & band\\ 
\hline                        
   0 & 1 - 0 & 43.42386 & 3.036$\times10^{-6}$ & 1 &  & 1 - 0 & 42.87992  & 2.114$\times10^{-8}$ & 1\\ 
     & 2 - 1 & 86.84699 & 2.915$\times10^{-5}$ & 2 &  & 2 - 1 & 85.75906 & 2.460$\times10^{-7}$  & 2\\\cline{1-5}
    & 1 - 0 & 43.12208  & 3.011$\times10^{-6}$ & 1 & 0 & 3 - 2 & 128.63685& 9.696$\times10^{-7}$ & 4\\
    & 2 - 1 & 86.24344  & 2.891$\times10^{-5}$ & 2 &  & 4 - 3 & 171.51255 & 2.532$\times10^{-6}$ & 5\\
    & 3 - 2 & 129.36326 & 1.045$\times10^{-4}$ & 4 &  & 5 - 4 & 214.38548 & 5.334$\times10^{-6}$ & 6\\
 1  & 4 - 3 & 172.48102 & 2.569$\times10^{-4}$ & 5 &  & 6 - 5 & 257.25493 & 9.869$\times10^{-6}$ & 6\\\cline{6-10}
    & 5 - 4 & 215.59592 & 5.131$\times10^{-4}$ & 6 &  & 3 - 2 & 127.74849 & 3.350$\times10^{-4}$ & 4\\
    & 6 - 5 & 258.70725 & 9.003$\times10^{-4}$ & 6 & 1 & 4 - 3 & 170.32807& 8.745$\times10^{-4}$ & 5\\
    & 7 - 6 & 301.81430 & 1.445$\times10^{-3}$ & 7 &  & 6 - 5 & 255.47849 & 3.407$\times10^{-3}$ & 6\\\cline{1-10} 
    & 1 - 0 & 42.82059  & 2.986$\times10^{-6}$ & 1 & 2 & 6 - 5 & 253.70317 & 1.11139 & 6\\\cline{6-10}
    & 2 - 1 & 85.64046  & 2.866$\times10^{-5}$ & 3 &  &  &      &  & \\
    & 3 - 2 & 128.45881 & 1.036$\times10^{-4}$ & 4 &   \multicolumn{5}{c}{$^{30}$SiO} \\\cline{6-10}
  2 & 4 - 3 & 171.27507 & 2.547$\times10^{-4}$ & 5 &   &  1 - 0  & 42.373426  & 2.016$\times10^{-8}$ &    1 \\
    & 5 - 4 & 214.08848 & 5.088$\times10^{-4}$ & 6 & 0 &  2 - 1  & 84.746170  & 2.346$\times10^{-7}$ &    2 \\
    & 6 - 5 & 256.89831 & 8.927$\times10^{-4}$ & 6 &   &  5 - 4  & 211.853473 & 5.081$\times10^{-6}$ &    6 \\\cline{6-10}
    & 7 - 6 & 299.70386 & 1.433$\times10^{-3}$ & 7 & 1 &  4 - 3  & 168.323352 & 8.054$\times10^{-4}$ &    5 \\\cline{1-10}
    & 1 - 0 & 42.51938  & 2.951$\times10^{-6}$ & 1 & 2 &  4 - 3  & 167.160563 & 2.542$\times10^{-1}$  &   5  \\\cline{6-10}
  3 & 3 - 2 & 127.55521 & 1.027$\times10^{-4}$ & 4 &  &          &            &  &      \\
    & 4 - 3 & 170.07057 & 2.525$\times10^{-4}$ & 5 &  &          &            &  &     \\
    & 5 - 4 & 212.58248 & 5.044$\times10^{-4}$ & 6 &   &         &            &  &      \\\cline{1-5}
  4 & 5 - 4 & 211.07784 & 4.986$\times10^{-4}$ & 6 &   &         &            &  &      \\
\hline\hline
\end{tabular}
\end{table*}

\begin{table*}
\caption{Einstein coefficient and frequencies of the detected H$_{2}$O and HCN maser transitions at submillimetre wavelengths.}             
\label{table:2}      
\centering          
\begin{tabular}{c c c c c || c c c c c}     
\hline\hline       
\multicolumn{5}{c}{H$_{2}$O} & \multicolumn{5}{c}{HCN} \\ 
\hline                    
     & Transition & Freq   &     A     & ALMA &      & Transition & Freq & A & ALMA\\
     &            & (GHz)  & (s$^{-1}$) & band &      &   & (GHz) & (s$^{-1}$) & band\\     
\hline
     & 3$_{13}$ - 2$_{20}$ & 183.31012 & 3.629$\times10^{-6}$  & 5 &    & 1 - 0 & 88.631602 & 1.771$\times10^{-7}$ & 2/3 \\ \cline{6-10}
     & 10$_{29}$ - 9$_{36}$ & 321.22564& 6.348$\times10^{-6}$ & 7 & $v_{2}=2^{0}$ & 1 - 0 & 89.0877 & 1.483$\times10^{-4}$ & 2/3\\ \cline{6-10}
     & 5$_{15}$ - 4$_{22}$  & 325.15292& 1.166$\times10^{-5}$ & 7 &    & 2 - 1 & 177.2387 & 4.578$\times10^{-5}$ & 5 \\                 
     & 17$_{413}$ - 16$_{710}$ & 354.8089 & 1.096$\times10^{-5}$ & 7 & $v_{2}=1^{1_{c}}$   & 3 - 2 & 267.1993 & 2.262$\times10^{-4}$ & 6\\
     & 7$_{53}$ - 6$_{60}$ & 437.34667 & 2.212$\times10^{-5}$ & 8 &   & 4 - 3 & 354.4605 & 6.122$\times10^{-4}$ & 7\\ \cline{6-10}                  
     & 6$_{43}$ - 5$_{50}$ & 439.15081 & 2.857$\times10^{-5}$ & 8 & $v_{2}=4$ & 9 - 8 & 804.7509 & - \tablefootmark{a} & 10 \\ \cline{6-10}
   
     & 6$_{42}$ - 5$_{51}$ & 470.88895 & 3.534$\times10^{-5}$ & 8 & $v_{1}=1^{1} \rightarrow v_{2}=4^{0}$  & 10 - 9 & 890.761& -\tablefootmark{a} & 10 \\\cline{1-5}
     & 4$_{40}$ - 5$_{33}$ & 96.26116 & 4.719$\times10^{-7}$ & 3  &  &  &  & & \\
$v_{2}=1$  & 5$_{50}$ - 6$_{43}$ & 232.68670 & 4.770$\times10^{-6}$ & 6 &  &  &  & &\\
     & 6$_{61}$ - 7$_{52}$ & 293.6645 & -\tablefootmark{a} & 6 &  & & & & \\
     & 1$_{10}$ - 1$_{01}$ & 658.00655 & 5.568$\times10^{-3}$ & 9 &  &  &  & & \\
\hline                  
\hline
\end{tabular}
\tablefoot{
\tablefoottext{a}{No spectroscopic data available}
}
\end{table*}

\section{Observations of (sub)millimetre masers}

\subsection{SiO}

Although it was first detected towards the Orion molecular cloud \citep{Snyder}, later surveys have
shown that SiO maser emission is quite uncommon in SFRs \citep{Zapata}. To date, the only submillimetre SiO masers detected in SFRs are
the $v=1$, $J=1-0$, $J=3-2$ and the $v=2$, $J=1-0$ rotational transitions \citep{Buhl, Davis}.
In contrast, strong SiO emission arises from the innermost regions of the CSE of late-type stars, and a number of maser transitions have been detected 
in vibrational levels of up to $v=4$, with rotational transitions as high as $J=8-7$ (\citealt{Menten}, \citealt{Humphreys} and references therein).
Interferometric obervations have revealed that the SiO maser emitting regions form ring-like structures centred
on the star, between the stellar photosphere and the dust-forming region $\sim 2$-$6$~R$_{\star}$ (\citealt{Diamond,Boboltz}). 
The most common SiO maser transitions detected are the two lowest rotational transitions of the $v=1$ vibrational level, 
peaking around 43~GHz and 86~GHz (e.g. \citealt{Barvainis}).
The list of SiO rotational transitions, including the isotopologues $^{29}$SiO and $^{30}$SiO, which can be observed as maser 
emission in the ALMA frequency range, are listed in Table~\ref{table:1} \citep{Mueller}.\\
Polarized SiO maser emission has been detected towards oxygen-rich (post-)AGB stars.
\citet{Barvainis} reported the detection of strong polarized emission in the rotational transitions $J=1-0$ and $J=2-1$, 
of both $v=1$ and $v=2$ vibrational levels.\\
Fractional linear polarization levels between 15$\%$ and 40$\%$ are commonly detected, though levels approaching 100$\%$ have also been 
reported (e.g. \citealt{Barvainis, Kemball, Amiri}). \citet{Shinnaga} and \citet{Vlemmings11}  
reported the detection of high fractional linear polarization levels ($p_{L}\geq$ 40\%) of the $J=5-4$ rotational transitions of the $v=1$ vibrational level, 
towards the supergiant VX Sgr. Such high polarization levels cannot only be explained by the presence of a large-scale magnetic field permeating 
the masing region, but need to be enhanced by non-Zeeman effects. 

\subsection{H$_{2}$O}

Water is among the most abundant molecules in the envelopes of cores embedded in star-forming regions and
in the CSEs of late-type stars (e.g. \citealt{Waters,Menten9a,Matthias}). At submillimetre wavelengths a number of molecular transitions of H$_{2}$O 
have been detected displaying high flux density values and narrow line shapes 
characteristic of maser emission lines \citep{Waters, Menten90, Yates}. 
The most studied water maser emission is the low-frequency $6_{16}$-$5_{23}$ transition at 22.2~GHz.
Since this transition is not affected by the atmospheric precipitable water vapor (PWV), it has become a reference in the study
of water maser emission from astrophysical sources. It has been detected towards the expanding CSEs of late-type stars and
high-velocity outflows generated in the envelope of both protostellar objects and (post-) AGB stars.  
This variety of scenarios gives us an idea of the broad excitation conditions of the 22.2~GHz maser 
(high-density regions n$_{H_{2}}>10^{8}$~cm$^{-3}$ and temperatures 2000~K $> T >$ 200~K, \citealt{Neufeld,Humphreys}).
The submillimetre water maser transitions within the ground-vibrational state have been
detected towards CSEs of O-rich late-type stars as well as from low- and high-mass SFRs, either 
tracing shocked regions or arising from the steadily expanding CSEs (e.g. \citealt{Ivison,Melnick}). The different water transitions observable in the ALMA 
frequency range within the ground-vibrational state and the $v_{2}=1$ bending mode are listed in the Table~\ref{table:2} \citep{Pickett}.
The rotational maser transitions within the vibrationally excited level $v_{2}=1$ are thought to arise from regions where the physical conditions are similar to 
those that invert the level population of the SiO rotational transitions \citep{Alcolea}.\\
In contrast to the case of SiO, there are not many references in the literature reporting the detection of H$_{2}$O maser polarized radiation in 
the ALMA frequency range. To date, only \citet{Harwit} have succeeded in measuring linear polarization 
of water maser radiation at 620.7~GHz. Unfortunately, the very low atmospheric transmission around 621 GHz prevents the
detection of this maser transition from ground-based telescopes.\\
The excitation conditions of the submillimetre water masers seem to be a subset of the broad excitation conditions 
generating the 22.2~GHz maser line \citep{Neufeld,Humphreys}. The dominating pumping process 
depends on the characteristics of the region where the maser emission arises; either from post-shock regions, 
where the inversion of the population of the molecular states is mainly collisional, or from the steadily expanding CSEs, 
where the pumping process is mainly thought to be caused by infrared photons of warm dust emission.  
Observations of 22.2~GHz H$_{2}$O maser polarized emission have been used to probe the magnetic field strengths within
the steadily expanding CSEs of AGB and supergiant stars (e.g.~\citealt{Vlemmings05}). In addition, polarized water maser emission has been detected
towards the high-velocity outflows of the so-called water-fountains, with spectral features displaying high levels of fractional linear 
polarization ($p_{L}> 5\%$) and unusually broad velocity ranges. These observations have probed magnetically collimated "jets" that appear
during the post-AGB phase, and are thought to be the precursors of bipolar (multipolar) planetary nebulae \citep{Vlemmings06, Perez-Sanchez}.   

\subsection{HCN}

Strong maser emission has also been detected from HCN in the CSEs of several carbon-rich (C-rich) AGB stars 
in the ALMA frequency range (e.g. \citealt{Bieging, Schilke}). 
Table~\ref{table:2} summarizes the different HCN transitions that can be observed as maser emission in the submillimetre 
wavelength regime \citep{Mueller}. Linearly polarized HCN maser emission of the 89.087~GHz, $J=1-0$ 
transition, within the (0,$2^{0}$,0) vibrationally excited state, has
been detected at approximately 20\% towards the innermost region of the CSE of the C-rich 
star CIT 6 \citep{Goldsmith}. The masing region is thought to be located between the photosphere and the inner radius of the expanding envelope, similar
to the SiO masers in the oxygen-rich late-type stars. 
The pumping of the HCN $J=1-0$ transition is more likely caused by the absorption 
of infrared photons, not by collisional processes, though a combination 
of both processes cannot be ruled out \citep{Goldsmith}. Anisotropies in the population inversion of the
masing transitions might affect the fractional linear polarization, but more observations are
needed to determine the exact cause of the polarization.

\section{Model results}

\begin{table}
\caption{Fixed parameter values assumed in our numerical models.}             
\label{table:t3}      
\centering                          
\begin{tabular}{c c c c c c}        
\hline\hline                 
Molecule & $g$ & $\Gamma$ & $g\Omega_{1\mathrm{G}}$\tablefootmark{a} & B\tablefootmark{b} & $\Delta\Omega$ \\    
 &  & (s$^{-1}$) & (s$^{-1}$) & (G) & (sr) \\
\hline\hline                        
  SiO & 0.155 & 5 & 1480.96  & 1 & $10^{-2}$ \\      
  H$_{2}$O & 0.65 & 1 & 6226.22  & 0.05 & $10^{-2}$ \\
  HCN  & 0.098 & 1 & 957.8 & 1 & $10^{-2}$ \\
\hline                                   
\end{tabular}
\tablefoot{
\tablefoottext{a}{Values calculated using Equation~\ref{eq:gomega} for a magnetic field $B=1$~G.}
\tablefoottext{b}{Magnetic field of the same order of magnitude as the values detected towards SFRs and CSE of late-type stars
for regions at densities of $n_{H_{2}}=10^{10}$~cm$^{-3}$ in the case of SiO and HCN, and $n_{H_{2}}=10^{8}$~cm$^{-3}$ in the case of H$_{2}$O.}
}
\end{table}

Maser emission can be affected by several non-Zeeman processes that can
enhance or even produce linear and/or circular polarization, such as anisotropic pumping
or the change of the quantization axis along the amplification path \citep{Wiebe}. Therefore, it is necessary to solve the radiative transfer equation
for a particular rotational transition to learn whether the polarization detected can be correlated to a large-scale magnetic field permeating the masing region. 
Furthermore, it is necessary to determine if the involved rotational transition can produce appreciable levels of $p_{L}$ when
the quantization axis is defined by the magnetic field direction, i.e. when the Zeeman frequency $g\Omega$ is higher
than the stimulated emission rate $R$.
The inequalities $g\Omega> \Gamma$ and $g\Omega> R$ allow us to analyse the polarization observed in terms of the ordinary population
of the magnetic substates \citep{Watson08}. In the case of SiO the decay rate $\Gamma$ listed in Table~\ref{table:t3} corresponds to the rate for the 
radiative decay from the first vibrationally excited state to the ground-vibrational state \citep{NW90}. For
water, this value is roughly the inverse
lifetime for infrared transitions of the 22~GHz, H$_{2}$O maser transition \citep{NW902}. In the case 
of HCN, this value corresponds to the decay rate associated to a pumping process dominated by infrared radiation \citep{G74}. 
Considering the $\Gamma$ and $g\Omega$ values in Table~\ref{table:t3}, it is clear that for the case of interest the criterium $g\Omega> \Gamma$ is satisfied. 
To evaluate whether the stimulated emission rate satisfies the condition $g\Omega> R$ for 
a particular transition when $p_{L}$ is sufficiently large to be detected, we have used a radiative transfer code adapted from 
\citet{NW92} and \citet{Vlemmings02}. The stimulated emission rate $R$ is given by

\begin{equation}
R\approx \frac{AkT_{b}\Delta\Omega}{4\pi hv},
\label{eq:equat}
\end{equation}

where $A$ is the Einstein coefficient of the involved transition, which is calculated and listed for most of the transitions
in Tables~\ref{table:1} and \ref{table:2}; $k$ and $h$ are the Boltzmann and Planck constant, $v$ is the maser frequency, $T_{b}$ is the brightness temperature 
and $\Delta\Omega$ is the relation between the real angular size of the of the masing cloud and the observed angular size. 
Although the quantities $T_{b}$ and $\Delta\Omega$ are related to the observed 
intensity of the maser, it is a difficult task to constrain their value directly from observations. 
Brightness temperatures of up to $\sim 10^{15}$~K have been measured for water masers in SFRs, whereas estimated 
values for $\Delta\Omega$ are $\sim 10^{-1}-10^{-2}$~sr for water masers detected towards the CSEs of late-type stars (\citealp{Anita11,Vlemmings051}), and 
in some cases $\Delta\Omega \sim 10^{-5}$~sr for the masers detected towards SFRs \citep{NW91}.\\
The fractional linear polarization increases with $T_{b}\Delta\Omega$ in the presence of a magnetic field as long as $\theta\neq \theta_{cr}$. 
Since $R$ increases faster than $p_{L}$ when $T_{b}\Delta\Omega$ increases, 
the polarization vector of linearly polarized maser radiation detected with high brightness temperature 
does not necessarily satisfy the criterium $g\Omega > R$, and
thus cannot always be directly correlated with the direction of the magnetic field lines, 
unless the emitted radiation has a high degree of beaming.
On the other hand, low brightness temperatures might result in very low or undetectable values of 
fractional linear polarization. Furthermore, the value of $T_{b}\Delta\Omega$ scales with the ratio $R/\Gamma$,
and higher values of the molecular decay rate $\Gamma$ imply a lower level of fractional linear polarization for
the same $T_{b}\Delta\Omega$ value. This reflects the dependence of the fractional linear polarization level of a particular 
rotational transition on the saturation level of the maser radiation.\\
The Zeeman frequency $g\Omega$ determines the energy-splitting of the magnetic sub-levels. 
In a masing region permeated by a constant magnetic field parallel to the z-axis of the coordinate system, 
the energy-shifting of the magnetic sub-levels is given by $\hbar g\Omega m/2 = g\mu_{N} B m$, where $\mu_{N}$ and $m$ are the nuclear magneton and
the quantum number of the magnetic substate \citep{NW90}. Hence the values for $g\Omega$ can be estimated 
using the relation

\begin{equation}
g\Omega= \frac{2g\mu_{N}}{\hbar} \frac{B\mathrm{[G]}}{1\times10^{4}},
\label{eq:gomega}
\end{equation}

where the factor $1\times10^{4}$ originates from the conversion between the units of Tesla and Gauss. In general, the molecular Land\'e factor ($g$) is different for each magnetic sub-level. 
The data in the literature are very limited, and there are no reported $g$-factor values for all the different rotational transitions of HCN nor H$_{2}$O. 
The Land\'e $g$-factor for SiO has minor differences (less than 1\%) for the $v=0,v=1$ and $v=2$ vibrational transitions 
\citep{Landolt}. The assumed $g$-factors for the different molecules are listed in Table~\ref{table:t3}.
In the case of H$_{2}$O, \cite{NW92} calculated the $g\Omega$ values considering hyperfine splitting for the $6_{16}$ and $5_{23}$ 
rotational states of water. Nevertheless, in the present case, we have assume the lower limit suggested for $g$ as 
it would be without hyperfine splitting \citep{Kukolich}, to calculate $g\Omega$ for the different H$_{2}$O rotational transitions that we modelled. 
For HCN we assumed the lower $g$ factor value reported for the (0,$1^{1_{c}}$,0) vibrational state, which corresponds to a 
magnetic field parallel to the molecular symmetry axis (\citealt{Goldsmith} and references therein). 
Because for closed-shell, non-paramagnetic molecules the response to a magnetic field is weak, 
we consider the $g$-factor values listed in Table~\ref{table:t3} as a conservative choice to constrain a minimum value for 
$g\Omega$ for the different molecular species. The $g\Omega$ values presented in Table~\ref{table:t3} correspond to
B$=1$~G in Equation~\ref{eq:gomega}, and can be scaled according to magnetic field strengths reported for the different
masing regions in both SFRs and CSEs of (post-)AGB stars.\\
Single-dish observations have revealed average magnetic field strengths of $\sim 3.5$ G in the SiO maser region, whereas
for H$_{2}$O the measured values range between 100$-$300~mG for the CSEs of AGB and supergiant stars, and
between 15$-$150~mG at densities of $n_{H_{2}}=10^{8}-10^{11}$~cm$^{-3}$ in SFRs. For our models we assumed
the values of the magnetic field strength listed in Table~\ref{table:t3}. This allows us to analyse our results
as a function of the minimum $g\Omega$ values for the three molecular species modelled.
Thus, using Equation~\ref{eq:gomega} for the corresponding magnetic field in Table~\ref{table:t3}, we calculate
the $g\Omega$ in order to determine whether the inequality $g\Omega > R$ is satisfied or not.

\section{Analysis}

\begin{table}
\caption{Model results: Values of $p_{L}$ assuming $\theta=90^{o}$ and $g\Omega = 10 R$ for rotational transitions of the $v=1$ vibrational state of SiO and 
the ground-vibrational level of H$_{2}$O. For HCN, the $J=1-0$ belongs to the vibrationally excited $v_{2}=2^{o}$ state, whereas the higher $J$ transitions 
listed are from the $v_{2}=1^{1_{c}}$ vibrationally excited level}             
\label{table:4}      
\centering                          
\begin{tabular}{c c || c c || c c}        
\hline
\multicolumn{2}{c}{SiO} & \multicolumn{2}{c}{H$_{2}$O} & \multicolumn{2}{c}{HCN} \\ 
\hline   
$J_{u}-J_{d}$ & $p_{L}$ & $J_{u}-J_{d}$ & $p_{L}$ & $J_{u}-J_{d}$ & $p_{L}$\\    
\hline                        
  1$-$0 & $\sim$0.32 & 3$_{13}$ - 2$_{20}$   & $\sim$0.19 & 1$-$0 & $\sim$0.33\\      
  2$-$1 & $\sim$0.23 & 5$_{15}$ - 4$_{22}$   & $\sim$0.16 & 2$-$1 & $\sim$0.27\\
  3$-$2 & $\sim$0.18 & 6$_{43}$ - 5$_{50}$   & $\sim$0.15 & 3$-$2 & $\sim$0.22\\ 
  4$-$3 & $\sim$0.15 &                     &            & 4$-$3 & $\sim$0.19\\
  5$-$4 & $\sim$0.14 & & \\
  6$-$5 & $\sim$0.13 & & \\
\hline
\hline
\end{tabular}
\end{table}

\subsection{SiO maser}
\label{sec:SiOR}
\begin{figure}
\centering
\includegraphics[width=9cm]{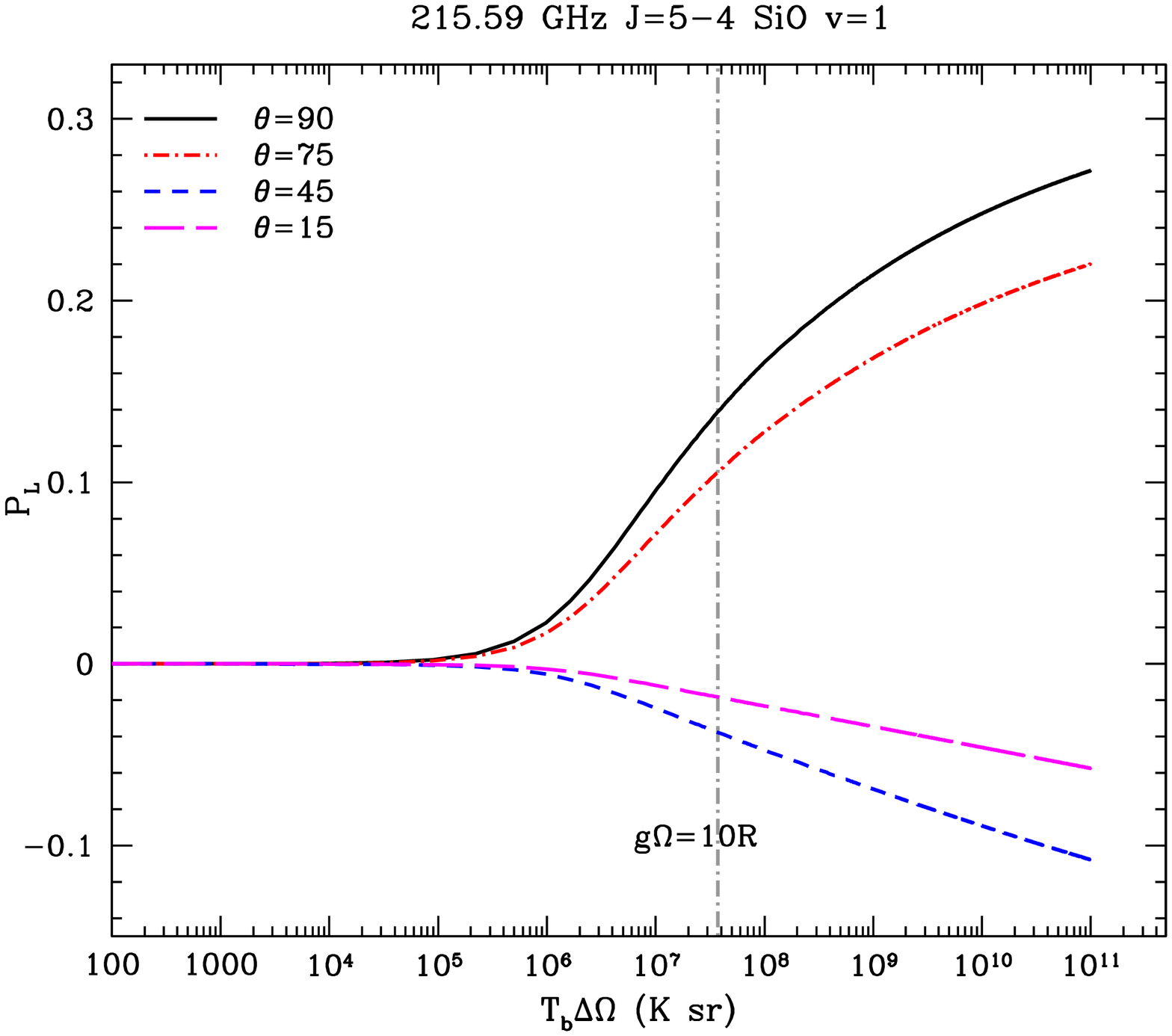}
\includegraphics[width=9cm]{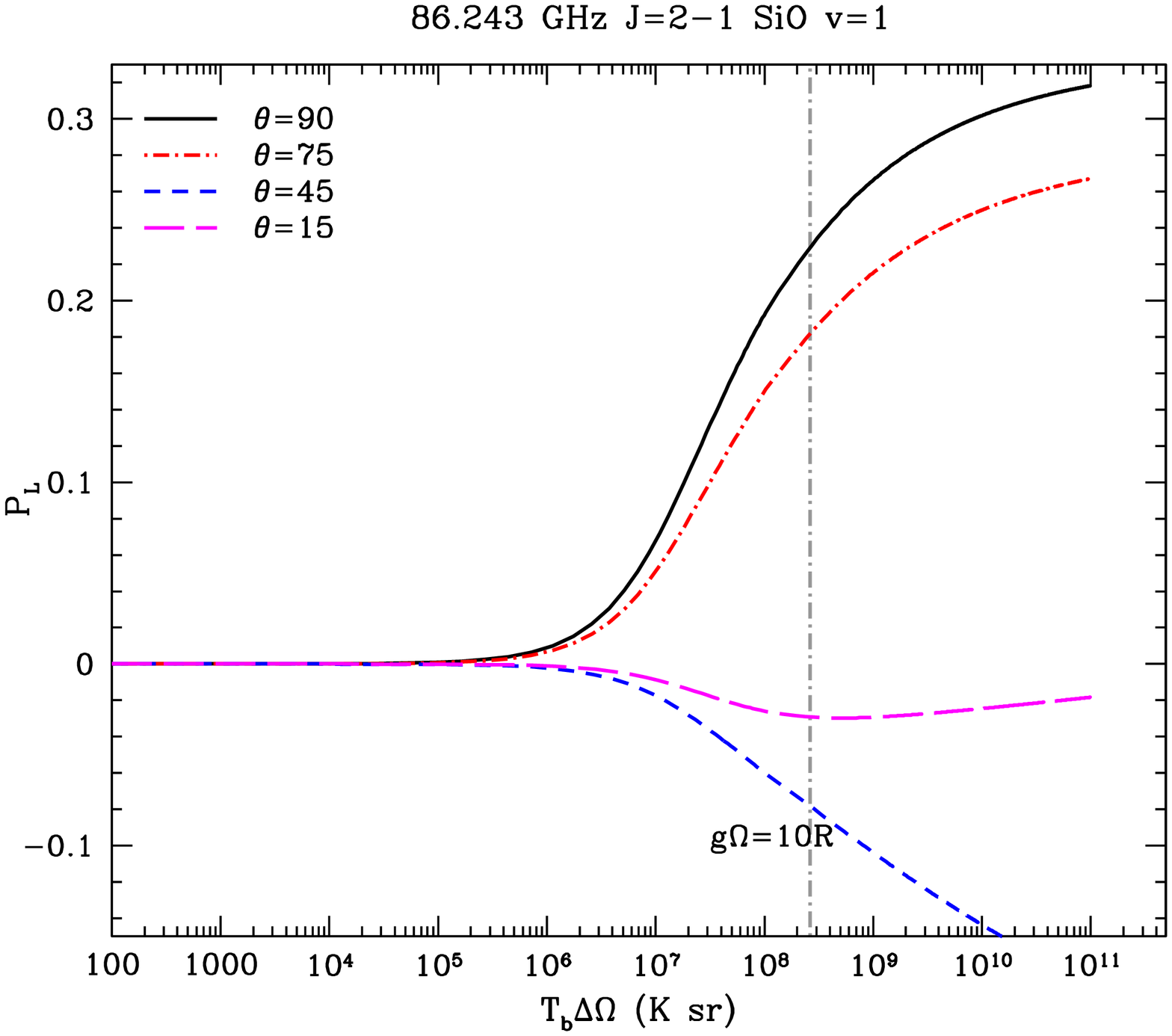}
\caption{Fractional linear polarization as a function of $T_{b}\Delta\Omega$ for the 215~GHz, $J=5-4$ (top), and the 86.2~GHz, $J=2-1$ (bottom) 
SiO rotational transitions within the $v=1$ vibrational level for four values of the angle between the magnetic field lines and the direction
of propagation of the maser radition, $\theta$, using the parameters listed in table \ref{table:t3} for SiO. Positive and negative $p_{L}$ values mean that the vector of polarization is either
perpendicular or parallel the magnetic field lines projected on the plane of the sky. The vertical line sets the $T_{b}\Delta\Omega$
value such that $g\Omega=10 R$.}
\label{Fig:oneSiO}
\end{figure}

We present the results of our models for the SiO rotational transitions $J=5-4$ and $J=2-1$ of the $v=1$ vibrational
state in Figure \ref{Fig:oneSiO}. The results are for four different
$\theta$ values and the vertical line corresponds to a $T_{b}\Delta\Omega$ where $g\Omega = 10 R$. The ideal case of $\theta=90^{o}$ determines
the maximum fractional linear polarization value when a magnetic field permeates the masing region.
The maximum fractional linear polarization for rotational transitions of the first SiO vibrationally excited state
are listed in Table~\ref{table:4}. These upper limits were established using the corresponding $g\Omega$ value listed in 
Table~\ref{table:t3} for each molecular species, and subsequently finding the $T_{b}\Delta\Omega$ 
values where $g\Omega = 10 R$. According to our results, it is possible to generate $p_{L}$ values of up to $13\%$ for the 
$J=5-4$ rotational transition, and of up to $23\%$ for the $J=2-1$ rotational transition without consideration of non-Zeeman effects 
(Figure \ref{Fig:oneSiO}), while the much higher values of $p_{L}$ detected for the SiO masers require anisotropic pumping.\\  
Therefore, although anisotropic pumping has a strong impact on the fractional linear polarization level, submillimetre SiO maser features 
observed with brightness temperatures $<10^{9}-10^{10}$~K fulfil $g\Omega > R$ and could in principle be used to trace the structure of the magnetic field permeating the SiO masing region.

\subsection{H$_{2}$O maser}

\begin{figure}
\centering
\includegraphics[width=9cm]{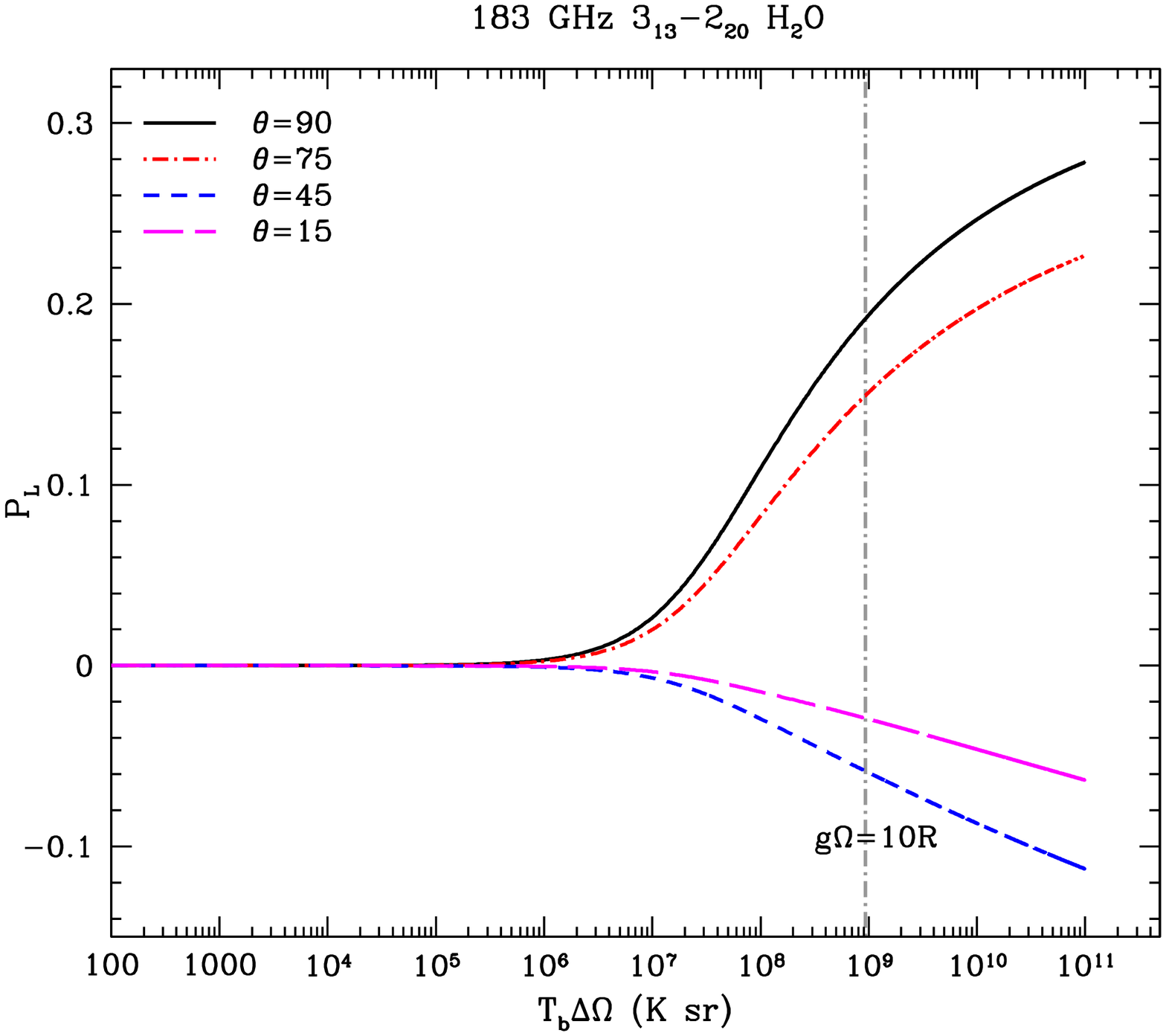}
\includegraphics[width=9cm]{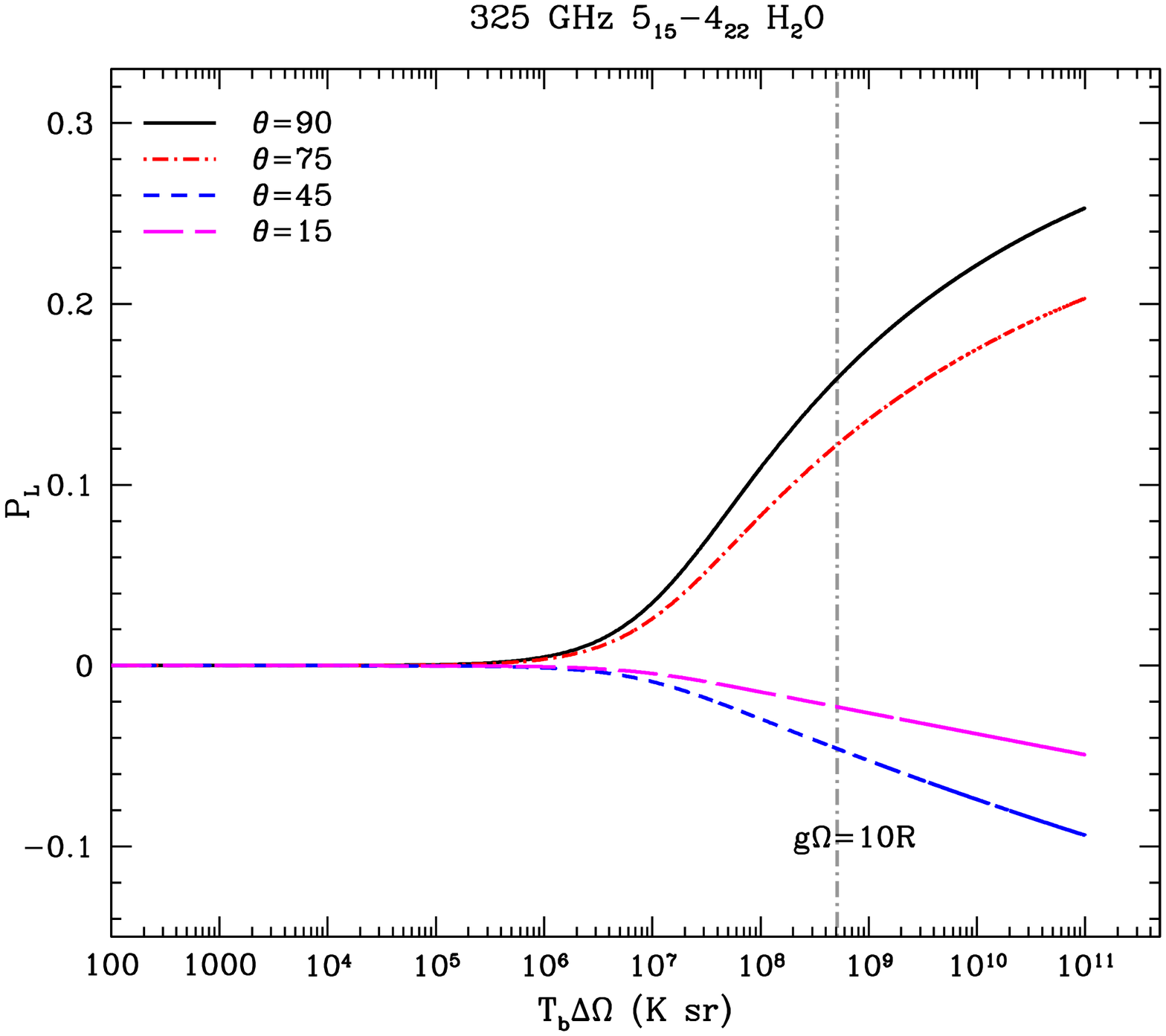}
\caption{Fractional linear polarization as a function of $T_{b}\Delta\Omega$ for the 183~GHz, $3_{13}-2_{20}$ (top), and the 325~GHz, $5_{15}-4_{22}$ (bottom) 
H$_{2}$O rotational transitions, both within the ground-vibrational level, for four values of the angle between the magnetic field lines and the direction
of propagation of the maser radiation, $\theta$. We considered the magnetic field strength and decay rate listed in table \ref{table:t3} for H$_{2}$O. 
Positive and negative $p_{L}$ values mean that the vector of polarization is either perpendicular or parallel the magnetic field lines projected on the plane of the sky. 
The vertical line sets the $T_{b}\Delta\Omega$ value such that $g\Omega=10 R$.}
\label{Fig:oneH2O}
\end{figure}

To determine the level of fractional linear polarization that water maser transitions can reach when the masing region
is permeated by a large-scale magnetic field, we ran numerical models of the rotational transitions within the
ground-vibrational state, using the assumptions described in section \ref{sec:SiOR}, together with the corresponding
parameters in table \ref{table:t3}. 
The results are presented in Table~\ref{table:4}.  In Fig.~\ref{Fig:oneH2O} we present the results for the 183~GHz 
and 325~GHz lines for four different $\theta$ values. Our results suggest upper limits for fractional linear polarization of up to 
$19\%$ and $16\%$ for the 183~GHz and 325~GHz lines,
respectively. Therefore, considering the values we assumed for the input parameters
listed in Table~\ref{table:t3}, maser features with $10^{8}$~K$< T_{b}< 10^{11}$~K can produce observable fractional linear polarization
levels that can be used to determine the magnetic field morphology. However, if $T_{b}$ is larger, the
observed polarization direction will be rotated away from the projected
magnetic field direction on the plane of the sky.

\subsection{HCN maser}

\begin{figure}
\centering
\includegraphics[width=9cm]{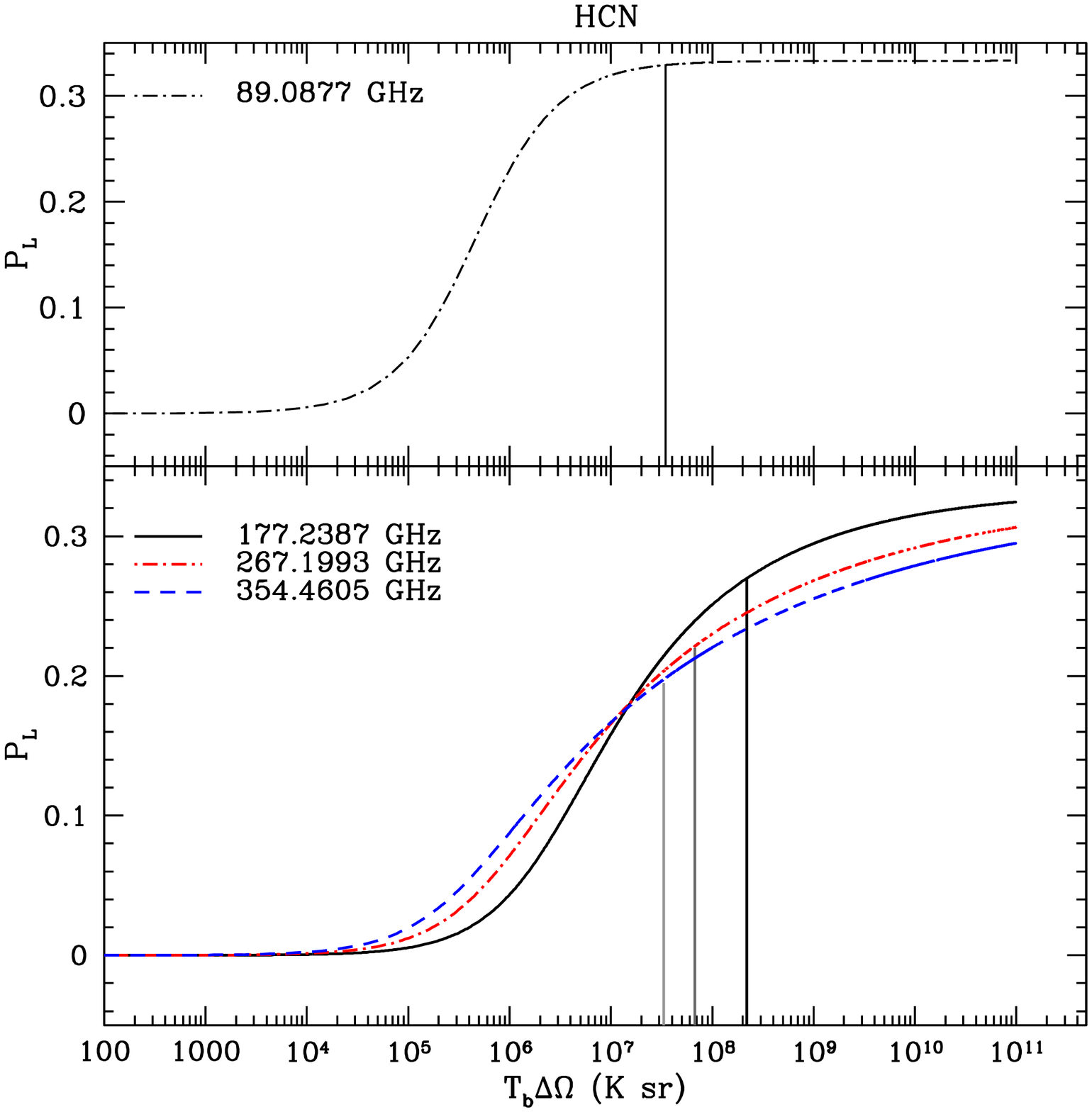}
\caption{Fractional linear polarization as a function of $T_{b}\Delta\Omega$ for $J=1-0$, $v_{1}=2^{0}$ rotational transition (top),
and three HCN rotational transitions of the vibrationally excited state $v_{1}=1^{1_{c}}$ (bottom), assuming a magnetic field oriented 
perpendicular to the direction of propagation of the maser radiation. We considered the magnetic field strength and decay rate listed in table \ref{table:t3} for HCN. 
Under these assumptions, the vertical lines set the maximum linear polarization that the different maser lines can reach 
satisfying the condition $g\Omega = 10 R$.}
\label{Fig:HCN}
\end{figure}

In Table~\ref{table:4} we present our results for the maximum fractional linear polarization that the $v_{2}=2^{0}$, $J=1-0$ rotational transition
and higher $J$ transitions within the $v_{2}=1^{1_{c}}$ vibrational state can produce. In Figure \ref{Fig:HCN} we show our results assuming
a magnetic field oriented perpedicular to the direction of propagation of the maser radiation, and the parameters listed in table \ref{table:t3}
for HCN. Fractional linear polarization of up to $\sim 33\%$ can be generated in the case of the $J=1-0$ rotational transition for $g\Omega= 10R$. This value 
decreases for rotational transitions with higher $J$ values, as expected. The values of $T_{b}\Omega$ scale with $R/\Gamma$, and consequently, 
lower $\Gamma$ values will increase the fractional polarization level that a particular rotational transition can reach while still satisfying 
$g\Omega> R$. Nevertheless, according to our results, HCN maser emission towards C-rich AGB stars can be generated with high fractional 
polarization values within the $g\Omega > R$ regime. Thus, the vector of polarization detected from HCN maser observations could be used to probe 
the magnetic field structure around C-rich AGB stars, even if the level of polarization has been enhanced by non-Zeeman effects.

\section{Observing maser polarization with ALMA}

\begin{table}
\caption{Sensitivity values at the different maser frequencies of SiO, H$_{2}$O and HCN within the ALMA frequency
range. The $\sigma$ listed corresponds to the sensitivity achieved after 1h of on-source time, with 0.1~km/s for a source
that reaches a maximum elevation of $59^{\circ}$.}             
\label{table:5}      
\centering                          
\begin{tabular}{c c || c c || c c}        
\hline
\multicolumn{2}{c}{SiO} & \multicolumn{2}{c}{H$_{2}$O} & \multicolumn{2}{c}{HCN} \\ 
\hline   
$v$ & $\sigma$ & $v$ & $\sigma$ & $v$ & $\sigma$\\    
(GHz) & (mJy) & (GHz) & (mJy) & (GHz) & (mJy)\\
\hline                        
86.243      & 4.58 & 183.310   & 38.37 & 89.0877  & 4.49\\      
129.363     & 4.32 & 325.153   & 39.58 & 177.238  & 4.97\\
172.481     & 4.74 & 439.151   & 59.25 & 267.199  & 4.12\\ 
215.596     & 3.63 &           &       & 354.461  & 4.99\\
258.707     & 4.76 & & \\
\hline
\hline
\end{tabular}
\end{table}

The temperature and density conditions that favour the population inversion of different
masing molecules in SFRs and CSEs of (post-) AGB stars are not the same. Most problably these conditions also differ
for different rotational transitions. Therefore, detecting polarized maser emission from multiple rotational transitions 
and from multiple molecular species with ALMA will enable us to trace the magnetic field structure throughout extended regions around
those sources. The accuracy of the ALMA polarimetry will allow us to detect polarized emission of 0.1~\% of the detected Stokes I.
Therefore, fractional linear polarization can be detected in short observations with very good 
signal-to-noise ratios. To probe whether the measured polarization of a maser traces
the magnetic field permeating the masing region, it is necessary to calculate the brightness temperature
of the spectral feature and evaluate if the product $T_{b}\Delta\Omega$ satisfies the inequality $g\Omega> R$ for
the observed emission. For detected spectral features it is possible to constrain $T_{b}$ by considering the equation

\begin{equation}
\frac{T_{b}}{[\mathrm{K}]}=\frac{S(v)}{[\mathrm{Jy}]}\Big(\frac{\Sigma^{2}}{[\mathrm{mas}^{2}]}\Big)^{-1}\zeta_{v},
\label{eq:tbtb}
\end{equation}  

where $S(v)$ is the detected flux density, $\Sigma$ is the maser angular size and $\zeta_{v}$ is a constant
factor that includes a proportionality factor obtained for a Gaussian shape \citep{Burns}. It scales with frequency according 
to the relation 

\begin{equation}
\zeta_{v}=6.1305\times 10^{11}\Big(\frac{v}{\mathrm{GHz}}\Big)^{-2} \frac{\mathrm{mas}^{2}}{\mathrm{Jy}}\mathrm{K}. 
\end{equation}

We calculated the rms value at the different maser frequencies that we modelled using the ALMA sensitivity calculator for an array of 
50 12-m antennas. The results are presented in Table~\ref{table:5} and correspond to observations
of 1~h of on-source time with 0.1~km/s of spectral resolution. 
To estimate the brightness temperature from observations that do not resolve individual features, it is necessary to assume 
a value for the size of the masing region generating the spectral feature. VLBI studies have revealed masers to be very compact 
spots with typical sizes of 1~AU. Based on observed peak-flux density values for maser emission of SiO, H$_{2}$O and HCN, assuming a size
of 1~mas, i.e. a source distance of 1~kpc, we here give a few examples of potential ALMA observations of maser polarization.\\
\\
{\it SiO masers:} SiO maser emission has been reported displaying flux density values between 7.4~Jy and 64~Jy \citep{Shinnaga}
for the $v=1$, $J=5-4$ rotational transition. Moreover, \cite{Barvainis} reported flux density values between 210~Jy and 625~Jy 
for the SiO $J=2-1$ rotational transition of the same vibrationally excited state. As an example,
we investigate the detection of a maser spectral feature with a peak-flux density of 10~Jy of
the SiO $J=5-4$ ($v=1$) rotational transition. For these values, Equation~\ref{eq:tbtb} gives $T_{b}=1.32\times10^{8}\ $~K. 
Comparing this with the $T_{b}\Delta\Omega$ value, which corresponds to the limit of $g\Omega= 10R$ (Figure \ref{Fig:oneSiO}), maser emission with
beaming angles $\Delta\Omega<2.84\times10^{-1}$~sr will have a linear polarization that can be used to trace the magnetic
field direction. But usually SiO maser spectra exhibit blended components. If this is the case, the flux density observed corresponds to a number 
of maser spots with similar line-of-sight velocities increasing the observed $S(v)$. Therefore, the brightness temperature derived 
using Equation~\ref{eq:tbtb} could be overstimated, but still could be
used to set an upper limit value for $T_{b}$. To reach the regime where $g\Omega > R$, it is necessary to associate a $\Delta\Omega$ value to 
the brightness temperature derived from the $S(v)$ of the blended components. Hence, since individual maser spectral features might have 
lower $T_{b}$ than the blended feature which contains it, even unbeamed maser radiation could still
place the linear polarization detected within the regime where $g\Omega > R$ is satisfied.\\
Finally, according to the ALMA sensitivity calculator, a $5\sigma$ detection of $1\%$ of fractional linear polarization of SiO maser emission
with low peak-flux densities (few tens of Jy/beam) requires short on-source observation time ($t < 1$~h).\\
\\
{\it H$_{2}$O masers:} Submillimetre H$_2$O lines have been detected with flux densities from a few tens to several thousands of Jy. 
For the minimun peak-flux density (243~Jy/beam) reported for the 183~GHz by \cite{Van}, the brightness temperature 
we obtain using Equation~\ref{eq:tbtb} is $T_{b}=4.43\times10^{9}$~K. Hence, H$_{2}$O maser
emission detected with beaming solid angles $\Delta\Omega\geq 1.6\times 10^{-3}$~sr can generate fractional linear polarization levels higher than 
$1\%$ (Figure \ref{Fig:oneH2O}), satisfying the condition $g\Omega > R$. The 1\% of linear polarization level of the weakest 183~GHz 
maser feature detected by \cite{Van} can be easily detected in very short integration times. Unfortunately, many H$_{2}$O maser lines are affected by 
the atmospheric PWV. For maser spectral features with peak-flux densities of the order of $20$~Jy/beam at 439.151~GHz, for instance, 
a $3\sigma$ detection of 1~$\%$ will require $\sim$ 1~h of on-source
observation time.\\ 
\\
{\it HCN masers:} HCN  $J=1-0$, ($v_{2}=2^{0}$) maser emission at 89.0877~GHz has been detected with fractional linear polarization of 20$\%$ 
for a spectral feature with flux density of 38~Jy \citep{Goldsmith}. Furthermore, \cite{Lucas} reported the detection of HCN $v_{2}=1^{1_{c}}$ maser 
emission with a peak-flux density $\sim$ 400~Jy at 177~GHz. \cite{Schilke20} reported the detection of the vibrationally excited HCN ($04^{0}0$) $J=9-8$
maser line near 805~GHz with a flux density of $\sim$1500~Jy.\\
For emission reported by \cite{Goldsmith}, assuming a magnetic field perpendicular to the propagation of the maser radiation, 
20\% of fractional linear polarization corresponds to $T_{b}\Delta\Omega=7.65\times10^{5}$~K~sr (Figure \ref{Fig:HCN}, top), which places the polarization
detected in the regime where $g\Omega > R $. On the other hand, considering a spectral feature of $S(v)=38$~Jy, Equation~\ref{eq:tbtb} 
gives $T_{b}=2.94\times 10^{9}$~K, implying a beaming angle of $ 2.6\times 10^{-4}$~sr, assuming a maser of 1~mas. A beaming angle 
$\Delta\Omega\sim 10^{-2}$~sr is needed for this maser to satisfy $g\Omega\geq 10R$ and be a realiable tracer of the magnetic field within the masing region.
However, if the emission consists of a contribution from multiple maser spots, the brightness temperature of the spectral feature will be 
overstimated, as discussed previously for blended SiO maser lines. Finally, as in the case of SiO maser emission, 1~\% of linear 
polarization could be observed with very low rms values in short on-source observation times (Table~\ref{table:5}). 

\section{Conclusions}

We have run numerical models to calculate the fractional linear polarization $p_{L}$
of maser emission generated by the interaction of a magnetic field with the different rotational transitions of 
SiO, H$_{2}$O and HCN within the ALMA frequency range. The fractional
linear polarization was calculated as a function of $T_{b}\Delta\Omega$, a quantity that can be related
to the stimulated emission rate $R$ of the involved transition. 
Considering both the minimun value of the Land\'e $g$-factor for each molecular species and a suitable magnetic 
field strength for the different masing regions, we found the maximum $p_{L}$ that the analysed rotational transitions can reach while 
satisfying both conditions $g\Omega > R$ and $g\Omega > \Gamma$. 
Meeting these criteria allow us to use the detected vector of polarization as a tracer of the magnetic field in the masing region,
even if the polarization observed has been affected by non-Zeeman effects.\\
According to our results, SiO, H$_{2}$O and HCN submillimetre maser emission can be detected with observable fractional polarization levels 
($>1\%$) in the regime $g\Omega > R$. But especially for the strongest masers, a careful analysis of the brightness temperature is needed to confirm
that the maser polarization is still in this regime. Thus, observing with ALMA full-polarization capabilities will enable us to 
use polarized maser emission as tracer of the magnetic field structure towards SFRs and CSEs of (post-)AGB stars. 
Depending on the maser spectral features detected, both the brightness temperature and the
beaming solid angle can be better constrained, leading to a more accurate determination of the direction of the 
magnetic field in the masing region. 
\\    
\begin{acknowledgement}
We wish to thank Matthias Maercker for his useful comments on the manuscript.
This research was supported by the Deutsche Forschungsgemeinschaft (DFG) through the Emmy Noether Research grant VL 61/3-1,
the DFG SFB 956 grant, and the BMBF ASTRONET project ARTIST.
\end{acknowledgement}

\bibliographystyle{aa} 
\bibliography{bibtex/refs} 

\begin{thebibliography}{63}
\expandafter\ifx\csname natexlab\endcsname\relax\def\natexlab#1{#1}\fi

\bibitem[{{Alcolea} \& {Menten}(1993)}]{Alcolea}
{Alcolea}, J. \& {Menten}, K.~M. 1993, in Lecture Notes in Physics, Berlin
  Springer Verlag, Vol. 412, Astrophysical Masers, ed. A.~W. {Clegg} \& G.~E.
  {Nedoluha}, 399--402

\bibitem[{{Alves} {et~al.}(2012){Alves}, {Vlemmings}, {Girart}, \&
  {Torrelles}}]{Alves}
{Alves}, F.~O., {Vlemmings}, W.~H.~T., {Girart}, J.~M., \& {Torrelles}, J.~M.
  2012, \aap, 542, A14

\bibitem[{{Amiri} {et~al.}(2010){Amiri}, {Vlemmings}, \& {van
  Langevelde}}]{Amiri10}
{Amiri}, N., {Vlemmings}, W., \& {van Langevelde}, H.~J. 2010, \aap, 509, A26

\bibitem[{{Amiri} {et~al.}(2012){Amiri}, {Vlemmings}, {Kemball}, \& {van
  Langevelde}}]{Amiri}
{Amiri}, N., {Vlemmings}, W.~H.~T., {Kemball}, A.~J., \& {van Langevelde},
  H.~J. 2012, \aap, 538, A136

\bibitem[{{Barvainis} \& {Predmore}(1985)}]{Barvainis}
{Barvainis}, R. \& {Predmore}, C.~R. 1985, \apj, 288, 694

\bibitem[{{Bieging}(2001)}]{Bieging}
{Bieging}, J.~H. 2001, \apjl, 549, L125

\bibitem[{{Boboltz} \& {Diamond}(2005)}]{Boboltz}
{Boboltz}, D.~A. \& {Diamond}, P.~J. 2005, \apj, 625, 978

\bibitem[{{Buhl} {et~al.}(1974){Buhl}, {Snyder}, {Lovas}, \& {Johnson}}]{Buhl}
{Buhl}, D., {Snyder}, L.~E., {Lovas}, F.~J., \& {Johnson}, D.~R. 1974, \apjl,
  192, L97

\bibitem[{{Burns} {et~al.}(1979){Burns}, {Owen}, \& {Rudnick}}]{Burns}
{Burns}, J.~O., {Owen}, F.~N., \& {Rudnick}, L. 1979, \aj, 84, 1683

\bibitem[{{Davis} {et~al.}(1974){Davis}, {Blair}, {van Till}, \&
  {Thaddeus}}]{Davis}
{Davis}, J.~H., {Blair}, G.~N., {van Till}, H., \& {Thaddeus}, P. 1974, \apjl,
  190, L117

\bibitem[{{Deguchi} \& {Watson}(1990)}]{Deguchi}
{Deguchi}, S. \& {Watson}, W.~D. 1990, \apj, 354, 649

\bibitem[{{Diamond} {et~al.}(1994){Diamond}, {Kemball}, {Junor}, {Zensus},
  {Benson}, \& {Dhawan}}]{Diamond}
{Diamond}, P.~J., {Kemball}, A.~J., {Junor}, W., {et~al.} 1994, \apjl, 430, L61

\bibitem[{{Dinh-v-Trung}(2009)}]{Dinh}
{Dinh-v-Trung}. 2009, \mnras, 399, 1495

\bibitem[{{Elitzur}(1992)}]{ElitzurB}
{Elitzur}, M., ed. 1992, Astronomical masers (Klwer Academic Publishers),
  Astrophys. Space Sci Lib, Vol. 170, {365}

\bibitem[{{Fish} \& {Reid}(2007{\natexlab{a}})}]{Fish07}
{Fish}, V.~L. \& {Reid}, M.~J. 2007{\natexlab{a}}, \apj, 656, 952

\bibitem[{{Fish} \& {Reid}(2007{\natexlab{b}})}]{Fish}
{Fish}, V.~L. \& {Reid}, M.~J. 2007{\natexlab{b}}, \apj, 670, 1159

\bibitem[{{Goldreich} {et~al.}(1973){Goldreich}, {Keeley}, \&
  {Kwan}}]{Goldreich}
{Goldreich}, P., {Keeley}, D.~A., \& {Kwan}, J.~Y. 1973, \apj, 179, 111

\bibitem[{{Goldreich} \& {Kwan}(1974)}]{G74}
{Goldreich}, P. \& {Kwan}, J. 1974, \apj, 190, 27

\bibitem[{{Goldsmith} {et~al.}(1988){Goldsmith}, {Lis}, {Omont}, {Guilloteau},
  \& {Lucas}}]{Goldsmith}
{Goldsmith}, P.~F., {Lis}, D.~C., {Omont}, A., {Guilloteau}, S., \& {Lucas}, R.
  1988, \apj, 333, 873

\bibitem[{{Gray}(2012)}]{grayB}
{Gray}, M., ed. 2012, Maser Sources in Astrophysics (Cambridge Astrophysics),
  Cambridge University Press, Vol.~1, {430}

\bibitem[{{Harwit} {et~al.}(2010){Harwit}, {Houde}, {Sonnentrucker}, {Boogert},
  {Cernicharo}, {De Beck}, {Decin}, {Henkel}, {Higgins}, {Jellema}, {Kraus},
  {McCoey}, {Melnick}, {Menten}, {Risacher}, {Teyssier}, {Vaillancourt},
  {Alcolea}, {Bujarrabal}, {Dominik}, {Justtanont}, {de Koter}, {Marston},
  {Olofsson}, {Planesas}, {Schmidt}, {Sch{\"o}ier}, {Szczerba}, \&
  {Waters}}]{Harwit}
{Harwit}, M., {Houde}, M., {Sonnentrucker}, P., {et~al.} 2010, \aap, 521, L51

\bibitem[{{Herpin} {et~al.}(2006){Herpin}, {Baudry}, {Thum}, {Morris}, \&
  {Wiesemeyer}}]{Herpin}
{Herpin}, F., {Baudry}, A., {Thum}, C., {Morris}, D., \& {Wiesemeyer}, H. 2006,
  \aap, 450, 667

\bibitem[{{Humphreys}(2007)}]{Humphreys}
{Humphreys}, E.~M.~L. 2007, in IAU Symposium, Vol. 242, IAU Symposium, ed.
  J.~M. {Chapman} \& W.~A. {Baan}, 471--480

\bibitem[{{Ivison} {et~al.}(1998){Ivison}, {Yates}, \& {Hall}}]{Ivison}
{Ivison}, R.~J., {Yates}, J.~A., \& {Hall}, P.~J. 1998, \mnras, 295, 813

\bibitem[{{Kemball} \& {Diamond}(1997)}]{Kemball}
{Kemball}, A.~J. \& {Diamond}, P.~J. 1997, \apjl, 481, L111

\bibitem[{{Kemball} {et~al.}(2009){Kemball}, {Diamond}, {Gonidakis}, {Mitra},
  {Yim}, {Pan}, \& {Chiang}}]{Kemball09}
{Kemball}, A.~J., {Diamond}, P.~J., {Gonidakis}, I., {et~al.} 2009, \apj, 698,
  1721

\bibitem[{{Kukolich}(1969)}]{Kukolich}
{Kukolich}, S.~G. 1969, \jcp, 50, 3751

\bibitem[{{Landolt-B\"ornstein}(1982)}]{Landolt}
{Landolt-B\"ornstein}. 1982, Group II, 14a, ed. K.-H. Hellwege and A.M.
  Hellwege (Berlin:Springer), p750

\bibitem[{{Leal-Ferreira} {et~al.}(2012){Leal-Ferreira}, {Vlemmings},
  {Diamond}, {Kemball}, {Amiri}, \& {Desmurs}}]{Ferreira}
{Leal-Ferreira}, M.~L., {Vlemmings}, W.~H.~T., {Diamond}, P.~J., {et~al.} 2012,
  \aap, 540, A42

\bibitem[{{Lucas} \& {Cernicharo}(1989)}]{Lucas}
{Lucas}, R. \& {Cernicharo}, J. 1989, \aap, 218, L20

\bibitem[{{Maercker} {et~al.}(2008){Maercker}, {Sch{\"o}ier}, {Olofsson},
  {Bergman}, \& {Ramstedt}}]{Matthias}
{Maercker}, M., {Sch{\"o}ier}, F.~L., {Olofsson}, H., {Bergman}, P., \&
  {Ramstedt}, S. 2008, \aap, 479, 779

\bibitem[{{Melnick} {et~al.}(1993){Melnick}, {Menten}, {Phillips}, \&
  {Hunter}}]{Melnick}
{Melnick}, G.~J., {Menten}, K.~M., {Phillips}, T.~G., \& {Hunter}, T. 1993,
  \apjl, 416, L37

\bibitem[{{Menten} {et~al.}(1990{\natexlab{a}}){Menten}, {Melnick}, \&
  {Phillips}}]{Menten9a}
{Menten}, K.~M., {Melnick}, G.~J., \& {Phillips}, T.~G. 1990{\natexlab{a}},
  \apjl, 350, L41

\bibitem[{{Menten} {et~al.}(1990{\natexlab{b}}){Menten}, {Melnick}, {Phillips},
  \& {Neufeld}}]{Menten90}
{Menten}, K.~M., {Melnick}, G.~J., {Phillips}, T.~G., \& {Neufeld}, D.~A.
  1990{\natexlab{b}}, \apjl, 363, L27

\bibitem[{{Menten} {et~al.}(2006){Menten}, {Philipp}, {G{\"u}sten}, {Alcolea},
  {Polehampton}, \& {Br{\"u}nken}}]{Menten}
{Menten}, K.~M., {Philipp}, S.~D., {G{\"u}sten}, R., {et~al.} 2006, \aap, 454,
  L107

\bibitem[{{M{\"u}ller} {et~al.}(2001){M{\"u}ller}, {Thorwirth}, {Roth}, \&
  {Winnewisser}}]{Mueller}
{M{\"u}ller}, H.~S.~P., {Thorwirth}, S., {Roth}, D.~A., \& {Winnewisser}, G.
  2001, \aap, 370, L49

\bibitem[{{Nedoluha} \& {Watson}(1990{\natexlab{a}})}]{NW90}
{Nedoluha}, G.~E. \& {Watson}, W.~D. 1990{\natexlab{a}}, \apj, 354, 660

\bibitem[{{Nedoluha} \& {Watson}(1990{\natexlab{b}})}]{NW902}
{Nedoluha}, G.~E. \& {Watson}, W.~D. 1990{\natexlab{b}}, \apjl, 361, L53

\bibitem[{{Nedoluha} \& {Watson}(1991)}]{NW91}
{Nedoluha}, G.~E. \& {Watson}, W.~D. 1991, \apjl, 367, L63

\bibitem[{{Nedoluha} \& {Watson}(1992)}]{NW92}
{Nedoluha}, G.~E. \& {Watson}, W.~D. 1992, \apj, 384, 185

\bibitem[{{Neufeld} \& {Melnick}(1991)}]{Neufeld}
{Neufeld}, D.~A. \& {Melnick}, G.~J. 1991, \apj, 368, 215

\bibitem[{{P{\'e}rez-S{\'a}nchez} {et~al.}(2011){P{\'e}rez-S{\'a}nchez},
  {Vlemmings}, \& {Chapman}}]{Perez-Sanchez}
{P{\'e}rez-S{\'a}nchez}, A.~F., {Vlemmings}, W.~H.~T., \& {Chapman}, J.~M.
  2011, \mnras, 418, 1402

\bibitem[{{Pickett} {et~al.}(1998){Pickett}, {Poynter}, {Cohen}, {Delitsky},
  {Pearson}, \& {M{\"u}ller}}]{Pickett}
{Pickett}, H.~M., {Poynter}, R.~L., {Cohen}, E.~A., {et~al.} 1998, \jqsrt, 60,
  883

\bibitem[{{Richards} {et~al.}(2011){Richards}, {Elitzur}, \& {Yates}}]{Anita11}
{Richards}, A.~M.~S., {Elitzur}, M., \& {Yates}, J.~A. 2011, \aap, 525, A56

\bibitem[{{Schilke} {et~al.}(2000){Schilke}, {Mehringer}, \&
  {Menten}}]{Schilke20}
{Schilke}, P., {Mehringer}, D.~M., \& {Menten}, K.~M. 2000, \apjl, 528, L37

\bibitem[{{Schilke} \& {Menten}(2003)}]{Schilke}
{Schilke}, P. \& {Menten}, K.~M. 2003, \apj, 583, 446

\bibitem[{{Shinnaga} {et~al.}(2004){Shinnaga}, {Moran}, {Young}, \&
  {Ho}}]{Shinnaga}
{Shinnaga}, H., {Moran}, J.~M., {Young}, K.~H., \& {Ho}, P.~T.~P. 2004, \apjl,
  616, L47

\bibitem[{{Snyder} \& {Buhl}(1974)}]{Snyder}
{Snyder}, L.~E. \& {Buhl}, D. 1974, \apjl, 189, L31

\bibitem[{{Surcis} {et~al.}(2011){Surcis}, {Vlemmings}, {Torres}, {van
  Langevelde}, \& {Hutawarakorn Kramer}}]{Surcis}
{Surcis}, G., {Vlemmings}, W.~H.~T., {Torres}, R.~M., {van Langevelde}, H.~J.,
  \& {Hutawarakorn Kramer}, B. 2011, \aap, 533, A47

\bibitem[{{van Kempen} {et~al.}(2009){van Kempen}, {Wilner}, \&
  {Gurwell}}]{Van}
{van Kempen}, T.~A., {Wilner}, D., \& {Gurwell}, M. 2009, \apjl, 706, L22

\bibitem[{{Vlemmings}(2002)}]{Vlemmings02}
{Vlemmings}, W. 2002, PhD thesis, , Leiden University, November 2002.

\bibitem[{{Vlemmings} {et~al.}(2006{\natexlab{a}}){Vlemmings}, {Diamond}, \&
  {Imai}}]{Vlemmings06}
{Vlemmings}, W.~H.~T., {Diamond}, P.~J., \& {Imai}, H. 2006{\natexlab{a}},
  \nat, 440, 58

\bibitem[{{Vlemmings} {et~al.}(2006{\natexlab{b}}){Vlemmings}, {Diamond}, {van
  Langevelde}, \& {Torrelles}}]{Vlemmings1}
{Vlemmings}, W.~H.~T., {Diamond}, P.~J., {van Langevelde}, H.~J., \&
  {Torrelles}, J.~M. 2006{\natexlab{b}}, \aap, 448, 597

\bibitem[{{Vlemmings} {et~al.}(2011){Vlemmings}, {Humphreys}, \&
  {Franco-Hern{\'a}ndez}}]{Vlemmings11}
{Vlemmings}, W.~H.~T., {Humphreys}, E.~M.~L., \& {Franco-Hern{\'a}ndez}, R.
  2011, \apj, 728, 149

\bibitem[{{Vlemmings} \& {van Langevelde}(2005)}]{Vlemmings051}
{Vlemmings}, W.~H.~T. \& {van Langevelde}, H.~J. 2005, \aap, 434, 1021

\bibitem[{{Vlemmings} {et~al.}(2005){Vlemmings}, {van Langevelde}, \&
  {Diamond}}]{Vlemmings05}
{Vlemmings}, W.~H.~T., {van Langevelde}, H.~J., \& {Diamond}, P.~J. 2005, \aap,
  434, 1029

\bibitem[{{Waters} {et~al.}(1980){Waters}, {Kakar}, {Kuiper}, {Roscoe},
  {Swanson}, {Rodriguez Kuiper}, {Kerr}, {Thaddeus}, \& {Gustincic}}]{Waters}
{Waters}, J.~W., {Kakar}, R.~K., {Kuiper}, T.~B.~H., {et~al.} 1980, \apj, 235,
  57

\bibitem[{{Watson}(2008)}]{Watson08}
{Watson}, W. 2008, in Cosmic Agitator: Magnetic Fields in the Galaxy

\bibitem[{{Watson}(2002)}]{Watson}
{Watson}, W.~D. 2002, in IAU Symposium, Vol. 206, Cosmic Masers: From
  Proto-Stars to Black Holes, ed. V.~{Migenes} \& M.~J. {Reid}, 464

\bibitem[{{Western} \& {Watson}(1984)}]{Western}
{Western}, L.~R. \& {Watson}, W.~D. 1984, \apj, 285, 158

\bibitem[{{Wiebe} \& {Watson}(1998)}]{Wiebe}
{Wiebe}, D.~S. \& {Watson}, W.~D. 1998, \apjl, 503, L71

\bibitem[{{Yates} \& {Cohen}(1996)}]{Yates}
{Yates}, J.~A. \& {Cohen}, R.~J. 1996, \mnras, 278, 655

\bibitem[{{Zapata} {et~al.}(2009){Zapata}, {Menten}, {Reid}, \&
  {Beuther}}]{Zapata}
{Zapata}, L.~A., {Menten}, K., {Reid}, M., \& {Beuther}, H. 2009, \apj, 691,
  332

\end{thebibliography}
\end{document}